# Function Follows Form: From Semiconducting to Metallic Towards Superconducting PbS Nanowires by Faceting the Crystal

Mohammad Mehdi Ramin Moayed[1,2], Sascha Kull[1], Angelique Rieckmann[1], Philip Beck[1], Michael Wagstaffe[3], Heshmat Noei[3], Andreas Kornowski[1], Ana B. Hungria[4], Rostyslav Lesyuk[1,5,6], Andreas Stierle[3,7], and Christian Klinke[1,5,8*]

[1] Institute of Physical Chemistry, University of Hamburg, Martin-Luther-King-Platz 6, 20146 Hamburg, Germany

[2] Deutsches Elektronen-Synchrotron DESY, Notkestrasse 85, 22607, Hamburg, Germany

[3] DESY NanoLab, Deutsches Elektronensynchrotron DESY, 22607 Hamburg, Germany

[4] Universidad de Cádiz, Facultad de Ciencias, Campus Rio San Pedro, Cadiz 11510, Spain

[5] Institute of Physics, University of Rostock, Albert-Einstein-Straße 23 - 24, 18059 Rostock, Germany

[6] Pidstryhach Institute for applied problems of mechanics and mathematics of NAS of Ukraine, Naukowa str. 3b, 79060 Lviv & Department of Photonics, Lviv Polytechnic National University, Bandery str. 12, 79000 Lviv, Ukraine

[7] Physics Department, University of Hamburg, 20355 Hamburg, Germany

[8] Department of Chemistry, Swansea University - Singleton Park, Swansea SA2 8PP, UK

*Correspondence to: christian.klinke@uni-rostock.de

**Abstract:** In the realm of colloidal nanostructures, with its immense capacity for shape and dimensionality control, the form is undoubtedly a driving factor for the tunability of optical and electrical properties in semiconducting or metallic materials. However, influencing the fundamental properties is still challenging and requires sophisticated surface or dimensionality manipulation. In this work, we present such a modification for the example of colloidal lead sulphide nanowires. We show that the electrical properties of lead sulphide nanostructures can be altered from semiconducting to metallic with indications of superconductivity, by exploiting the flexibility of the colloidal synthesis to sculpt the crystal and to form different surface facets. A particular morphology of lead sulphide nanowires has been synthesized through the formation of {111} surface facets, which shows metallic and superconducting properties in contrast to other forms of this semiconducting crystal, which contain other surface facets ({100} and {110}). This effect, which has been investigated with several experimental and theoretical approaches, is attributed to the presence of lead rich {111} facets. The insights promote new strategies for tuning the properties of crystals as well as new applications for lead sulphide nanostructures.

**Introduction**

In the recent decades it has been shown that the size reduction of different materials has led to the emergence of many novel properties due to the employment of quantum confinement[1]. For instance, the realization of spin-dependent phenomena[2,3], unconventional superconductivity regimes[4], and topological surface states[5] attracted considerable attention. Nevertheless, there is still room to tune the shape and size of nanomaterials, which might further modify their properties or even fundamentally change their character[6,7]. In this respect, the colloidal synthesis of nanomaterials could play a crucial role, since it shows a great degree of flexibility in tuning the product properties. Further, it is cheap, fast, and scalable, which makes it suitable for commercial applications[8,9]. The method has recently been employed to synthesize a variety of materials with different properties[6,8]. One example for the successful implementation of this approach is the synthesis of lead sulphide, which has been produced with a broad spectrum of shapes, sizes and properties[10-12]. This material has been already used for many applications such as photodetectors[13], field-effect transistors[14], spintronic components[2], and solar cells[15, 16]. Regarding all of these applications, a certain statement is valid: PbS exhibits semiconducting properties, which is not a surprising fact considering the electronic structure of this material[2, 6, 8, 10-12, 14-22]. However, violating this statement could be of great scientific and practical importance, since it establishes new strategies to tune the properties of crystalline materials based on their target applications.

Here, we introduce a method to change the electrical properties of colloidal lead sulphide nanowires from normal semiconducting to metallic with indications of superconductivity. This could be achieved by faceting the crystal, or in other words, by altering the surface facets of the crystal to the {111} ones, which are single element facets. This Pb-rich surface provides delocalized surface states at room temperature or presumably Cooper pairs at low temperatures, causing metallic and supposedly superconducting properties, in contrast to other forms of PbS nanocrystals, which are all semiconducting. Altering the surface facet is done by ligand mediated growth in the presence of oleic acid, lithium chloride and trioctylphosphine, with expressed {111} facets giving a zigzag shape.

Such zigzag wires are synthesized together with nanostripes, which have a flat shape, containing Pb and S atoms on their surface. Comparable to the earlier investigated PbS nanosheets[14], these straight nanostripes show semiconducting behaviour. By altering the synthesis conditions, especially the ligand combination, it is possible to produce predominantly metallic wires, semiconducting ones, or mixed products. Metallic behaviour of the zigzag wires is theoretically predicted by density-functional theory (DFT) calculations and experimentally confirmed by various transport measurements, including field effect, photoconductivity, and the temperature dependency of the conductivity. The origin of the effect is also investigated by high-resolution transmission-electron microscopy (HRTEM), scanning transmission-electron microscopy (STEM), X-ray photoelectron spectroscopy (XPS), and energy-dispersive X-ray spectroscopy in high angle annular dark field mode (HAADF-XEDS).

**Results**
**Synthesis and crystallography**
PbS nanowires with lengths of more than 10 µm and diameters of 30-50 nm have been synthesized according to the protocol presented in the Materials and Methods section. Based on the work of Bielewicz et al.[12], the morphology of these wires was influenced by employing oleic acid, TOP and halide ions as the ligand system, leading to the formation of a zigzag shape. Straight nanostripes could also be produced as the side product of this synthesis.

Such zigzag nanowires are supposed to be formed by oriented attachment of initial nuclei, lining up as thin wires with the diameter of less than 10 nm, and their further growth to the final structure (zigzag wires) (Supplementary Figure 1). The growth of these nanowires can be controlled by changing the composition of the ligand system, especially by the amount of oleic acid, to either suppress {111} faceting or to grow thicker zigzag wires. (Supplementary Figure 2). Halide ions on the other side are important to direct the oriented attachment to form 1D nanowires. Decreasing the concentration of halide ions promotes the 2D growth (formation of nanosheets), while increasing their amount hinders the anisotropic growth (formation of nanoparticles), analogue to the reports about the synthesis with halogenated hydrocarbons[12].

Although these nanostructures (straight stripes and zigzag wires) are the products of a single synthesis, their crystals exhibit important differences. As can be observed in Fig. 1a, the first distinction is their shape. While one type has a smooth and flat surface (Fig. 1b), the other type grows with zigzag shape and therefore, its surface is rough (Fig. 1c). 3D visualized STEM tomography of the zigzag wires and straight stripes can be found as Supplementary Videos. Small octahedra are the by-product of this synthesis and can be observed in the vicinity of these wires (Fig. 1a). The morphology of these crystals shows evident distinctions, revealed by HAADF-STEM. Regardless of the different appearances, the structures have the same Z contrast, since they are made of exactly the same material. Therefore, we assumed that the intensity of the HAADF signal is primarily determined by the thickness of the structures. As can be observed in Fig. 1d, the straight nanostructures (which are referred as nanostripes, since their width is higher than their thickness) are flat. This can be recognized by the uniform contrast distribution along the crystal. In contrast, the zigzag structures (which are referred as nanowires due to their comparable height and width) are uneven, which can be recognized by the undulating contrast distribution along the wire. Further, it can be seen that the flat stripes are thinner than the zigzag wires. The difference in shape of these nanostructures could be a sign for different growth and faceting mechanisms. Therefore, the samples have been investigated by high-resolution TEM to gain more details.

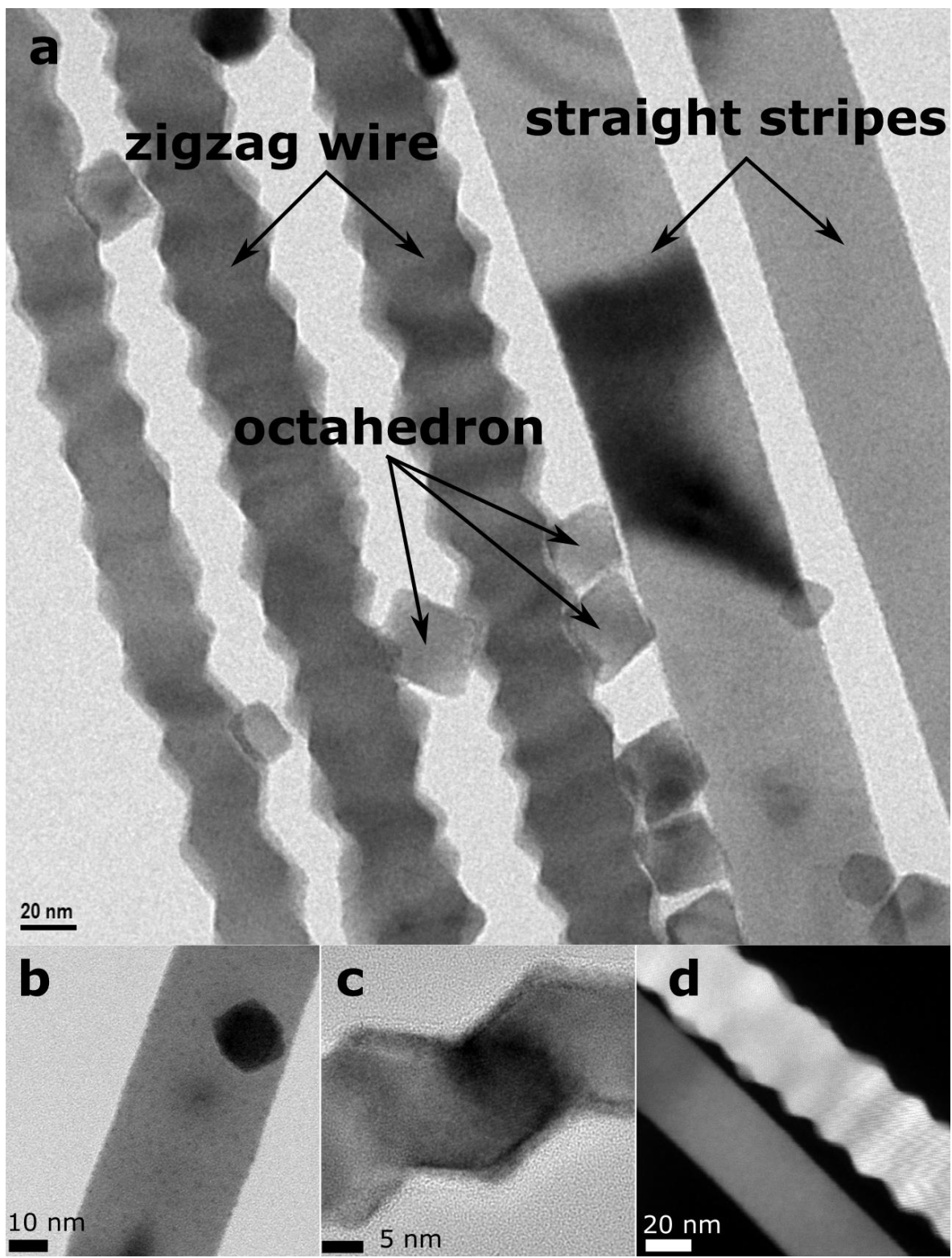

**Figure 1 | TEM images of the PbS nanostructures, grown with two different shapes.** (**a**) Two types of nanostructures are observable: straight stripes and zigzag wires. In addition, small octahedra are also present. (**b**) The first type grows in a straight form and has a quasi-flat surface, referred as straight nanostripes. (**c**) The second type has a zigzag form and therefore a corrugated surface, referred as zigzag nanowires. (**d**) Distinct morphology of the zigzag wires

and flat stripes, observable in HAADF-STEM image via the contrast difference. The straight nanostripes are flat and thinner than the uneven zigzag nanowires.

Figure 2a shows a high-resolution TEM image of a zigzag wire. These wires are observed along the <110> zone axis, with a <100> growth direction. These observations are in agreement with the literature[11, 17, 18, 23-25]. The atomic arrays on the surface of the zigzag wires form an angle of 109° (Fig. 2b). On the other hand, HRTEM images of straight nanostripes (Fig. 2c) show that their growth direction is the <112>. This is also in agreement with the growth direction of similar straight nanostripes, discussed in the literature[17].

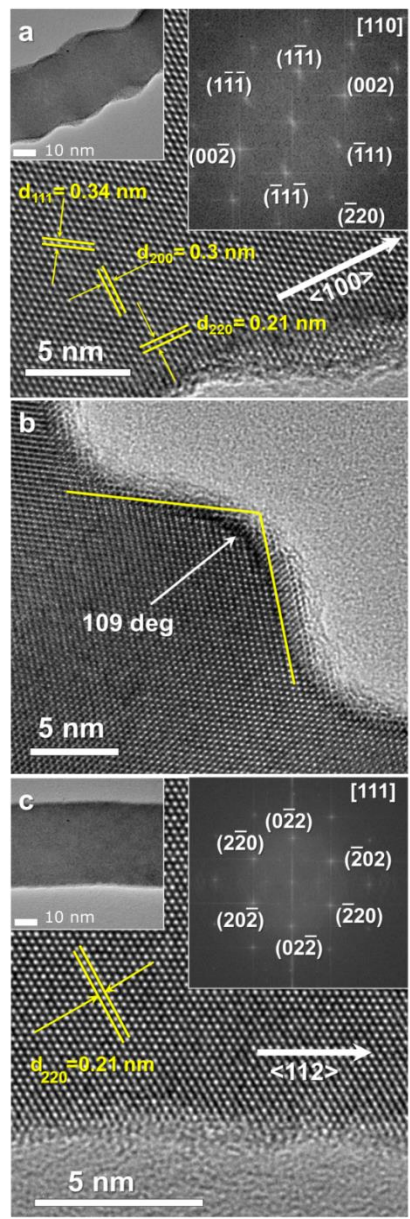

**Figure 2 | HRTEM images of the structures.** (**a**) Segment of a zigzag wire with the (200), (220), and (111) spacings. The direction of the growth was identified as <100>. Insets: Zoom out and corresponding Fast Fourier Transform (FFT) (zone axis <110>). (**b**) The angle between the atomic arrays at the surface of the zigzag wires is equal to 109°. (**c**) Segment of a straight

nanostripe with the (220) spacing. The growth direction is indexed as <112>. Insets: Zoom out and corresponding FFT (zone axis <111>).

Based on these observations, the faceting mechanism of the structures could be compared with the models suggested in the literature. During the formation of the straight nanostripes, first small polyhedra are formed. Then, they attach via the {112} facets and form stripes in the <112> direction. The faceting of the straight stripes is schematically shown in Fig. 3a. Since the building blocks of these stripes are comparable to nanosheets, their electrical properties are also similar (semiconducting)[12, 17].

On the other hand, the zigzag wires are supposedly formed through the attachment of octahedra. These octahedra are attached via the {100} facet (tip to tip) and form zigzag wires in the <100> direction[11, 17, 18, 23-25]. This type of attachment is shown in Fig. 3b. As can be seen, the top face of these wires is the {110} plane (in agreement with the HRTEM images), which in turn consists of {111} facets. In fact, this is the only stable way for them to lie on the TEM grid. Under these conditions, the surface atomic arrays must be observed with an angle of 109°, which is also true for the synthesized zigzag wires.

Formation, growth and ripening processes for the two types of nanostructures have different peculiarities. Zigzag wires are formed after the attachment of octahedral building blocks and grow further along the thermodynamically low-energy direction, i.e. <100>. The given synthesis conditions allow only the development of the {111} facets during the ripening phase. In contrast, straight nanostripes grow along the higher energy direction <112>, supposedly exposing the {100}/{110} facets with an envelope surface, which corresponds to the {111} planes as shown in Fig. 3a.

The existence of octahedra next to the zigzag wires (observable in the TEM images) also indicates the growth of small building blocks to octahedra (Fig. 1a), as we expect their growth to be similar to the zigzag wire growth (forming the {111} facets). Due to the extremely high Pb/S ratio in the synthesis (128:1), and also due to the high reactivity of the S rich {111} facets[19, 26], the surface of the zigzag wires is only composed of the Pb-rich {111} facets, as can be schematically seen in Fig. 3c. As will be shown later, the Pb-rich surface of these wires triggers some novel properties, including metallic behaviour.

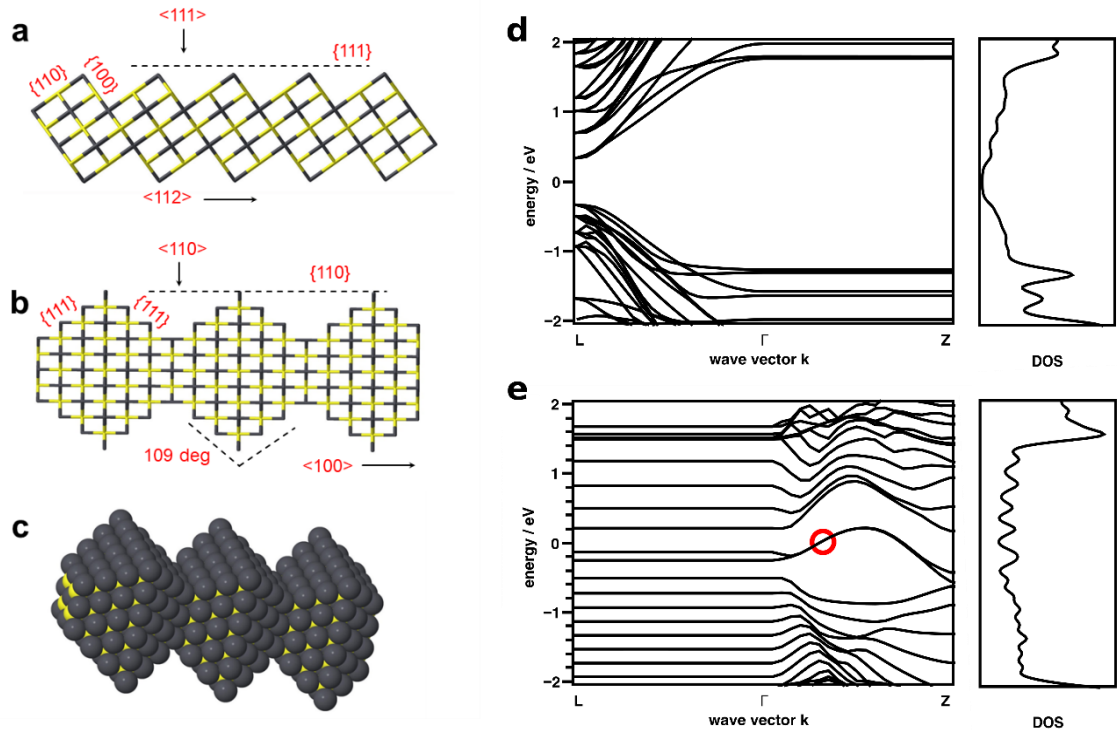

**Figure 3 | Faceting, growth direction, and electronic properties of different types of nanostructures.** (**a**) The straight nanostripes grow in the <112> direction via attachment of cubes via the {112} facets. Their top facet is viewed from the <111> direction. (**b**) The zigzag wires with the growth direction of <100> are made of octahedra, which are attached via the {100} facets. Their surface comprises of {111} facets viewed from <110> direction and the tips are observed with the angle of 109°. (**c**) Formation of the zigzag wires. The surface of these wires is composed of Pb atoms only, which are shown here as grey spheres (the yellow spheres represent the S atoms). (**d**) The band structure and the DOS for the crystal, cut along the {100} facet (representing the straight nanostripes), showing a DOS of zero in the bandgap. The band structure is flat on the path Γ-Z due to confinement in this direction. (**e**) The band structure and the DOS of a crystal cut along the {111} facet with Pb surface termination (representing the zigzag wires), which shows a throughout non-zero DOS. The band structure is flat on the path L-Γ due to confinement in this direction. The red circle marks one of the additional states at the Fermi level, which lead to metallic behaviour. It is visualized in the Supplementary Figure 3. In (d) and (e) the Fermi level is set to $E_F = 0.0$ eV.

**DFT calculations**

After identifying the faceting mechanism of the wires, their electrical properties can be predicted by DFT calculations. For this purpose, the band structure of the PbS crystal was calculated, on the one hand by cutting the crystal through the {100} facet and on the other hand by cutting through the {111} facet (without ligands). First, the bulk band structure has been verified using the unit cell and adequate cells ready for cutting along the mentioned facets under periodic boundary conditions. Then, 2D slabs have been tested with various thicknesses, surface terminations (Pb/Pb, S/S, Pb/S) and a vacuum section in the according thickness direction. The maximum thicknesses is 25 atomic layers in the shown case of {111} facets terminated with lead on both sides. In the case that at least one side was terminated by lead, the simulations led to metallic behaviour (additional bands at the Fermi level in the "bulk bandgap"). As can be seen in Fig. 3d, the density of states (DOS) of the crystals cut along the {100} surface becomes zero in a certain energy range, which shows the bandgap, which is slightly increased due to the confinement in one dimension, or in other words, the semiconducting character of this crystal. In contrast, for the case that the crystal is cut along the {111} facet terminated by lead on both sides, the band structure shows additional bands and the DOS is throughout non-zero (Fig. 3e), which yields a metallic behaviour for such crystals. Further, a visualization of one of the additional surface states, which can be found in Supplementary Figure 3, shows that the surface states has delocalized character, which is another indication for supposedly metallic behaviour. The conclusion of these calculations, which are in agreement with previously reported works[19, 27, 28], is that the crystal with the {100} cut (the straight nanostripes) must represent normal semiconducting behaviour of PbS but with the {111} cut (the zigzag wires), it must have a metallic character. An additional termination with a dense layer of Cl (every superficial sulphur or lead atom) led to more semiconducting behaviour, as discussed later.

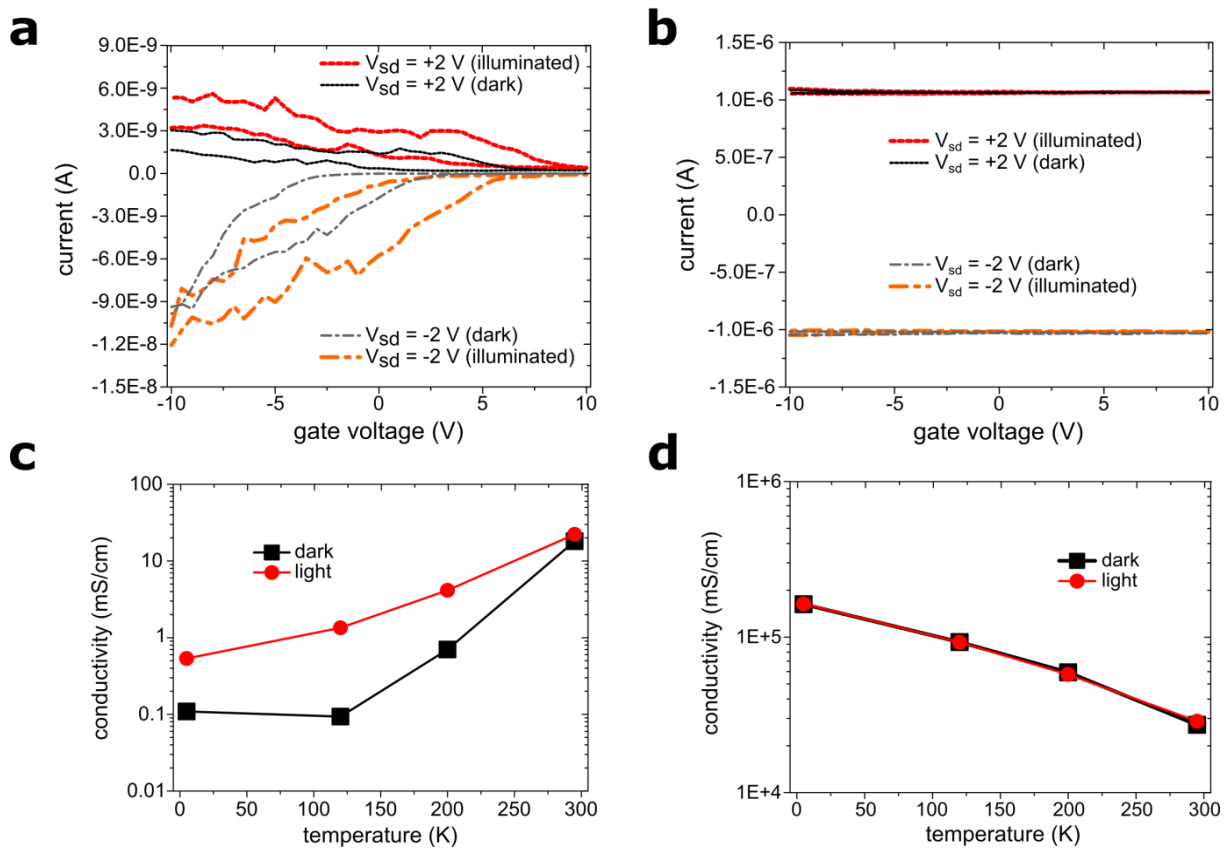

**Figure 4 | Electrical properties of the straight nanostripes and zigzag wires.** (**a**) The transfer characteristics of the straight nanostripes shows that their conductivity increases by applying a negative gate voltage or by laser illumination ($\lambda = 630$ nm), as expected for a semiconductor. (**b**) The transfer characteristics of the zigzag wires shows that the current, which is significantly higher than the current of the stripes, is adjustable neither by the gate nor

by illumination, representing metallic behaviour. (**c**) Temperature dependency of the conductivity and the photoconductivity for the straight nanostripes (gate voltage: 0 V). At low temperatures, the conductivity and the photoconductivity are lower, while the photocurrent to dark current ratio is higher. (**d**) The temperature dependency of the conductivity and the photoconductivity for the zigzag wires (gate voltage: 0 V). The conductivity increases by reducing the temperature. Photoresponse is absent at all temperatures.

**Transistor behaviour of the wires**

In order to experimentally investigate the differences between the electrical properties of these nanostructures, they have been contacted individually with Au electrodes and by means of electron-beam lithography. Supplementary Figure 4a demonstrates an exemplary scanning-electron microscope (SEM) image of the fabricated devices. The shape of the employed wires for each device has been determined by means of SEM images after performing the electrical measurements (the difference between these devices can be seen in Supplementary Figure 4b and c).

After the device fabrication, the sample has been immediately transferred to a vacuum probe station and measured with a back-gate device geometry. Figure 4a shows the results of the measurements on the straight nanostripes. Their transfer characteristics demonstrate the clear dependency of their conductivity to the gate voltage, showing p-type behaviour. The conductivity of these stripes is calculated to be 18 mS/cm. The gate dependency is the first sign for their semiconducting behaviour.

For the next step, the photoconductivity of these nanostructures has been investigated by illumination with a red laser ($\lambda$ = 630 nm). The straight nanostripes show an increase in the conductivity when they are illuminated with the laser, which is observable in the transfer characteristics (Fig. 4a). Illuminating these semiconducting crystals results in the optical excitation of the carriers over the bandgap, which increases the amount of the free charge carriers and proportionally enlarges the conductivity[20, 21].

Similar experiments have also been conducted on the zigzag wires. As can be seen in Fig. 4b, the transfer characteristics of the zigzag wires shows that their conductivity cannot be adjusted by the gate-electric field. Even by applying higher gate voltages (higher than 10 V), no current modulation could be observed. The conductivity of these wires reaches 27260 mS/cm, which is significantly higher than the conductivity of their semiconducting analogue. Further, no improvement of the conductivity is detectable when the zigzag wires are illuminated with the laser. As can be seen in Fig. 4b, the dark current (shown as black lines) always overlaps with the photocurrent (shown as red lines). Due to the metallic character of these wires and because of the overlap between their valence band and their conduction band, the amount of their mobile carriers is not affected by applying a gate-electric field or by the optical excitation[29]. The output characteristics of these devices can be found in Supplementary Figure 5.

**Temperature dependency**

To further investigate the properties of these nanostructures, the samples have been cooled down in order to observe the temperature dependency of the conductivity. Figure 4c demonstrates the conductivity and photoconductivity of the straight nanostripes at different temperatures, while the gate voltage is kept at 0 V (for more details see Supplementary Figure 6a). The conductivity of the stripes shows a positive temperature coefficient. By decreasing the temperature, the conductivity of the stripes decreases and therefore, a lower current flows through them. A similar behaviour has been already observed for PbS nanosheets[14]. The photoconductivity of these crystals also shows a similar trend (positive temperature coefficient), while the ratio of the photocurrent to the dark current increases at low temperatures. All of these observations are in agreement with the behaviour of semiconductors. At low temperatures, the thermal excitation of the carriers is suppressed and therefore, less mobile carriers can contribute to the transport[14].

In contrast, the temperature dependency of the zigzag wires, depicted in Fig. 4d, shows the increase of the current at lower temperatures (more details in Supplementary Figure 6b). The conductivity has a negative temperature coefficient and therefore, it is maximum at 5 K (the lowest measured temperature), while no photocurrents are detected even at low temperatures. These measurements are another proof for the metallic character of the zigzag wires. In metals, the carrier concertation is not governed by the thermal excitation. Instead, by reducing the temperature, phonon scattering is diminished, which increases the diffusion length of the carriers and therefore, the conductivity is improved[30].

**Four-point measurements**
In order to exclude the effect of the contact resistance, the zigzag wires, which show unusual metallic behaviour, have been measured with four-point geometry (shown in Supplementary Figure 7). As can be seen in the inset of Fig. 5a, by excluding the effect of the contact resistance, a linear I-V characteristics is observed. It indicates that the contact resistance is responsible for the S-like shape of the wires' output characteristics (Supplementary Figure 5b and 6b). The room temperature conductivity of the crystal (without the contact resistance) is calculated to be 99186 mS/cm, which is comparable or even higher than the reported values for bulk galena[31]. By decreasing the temperature, a clear reduction is detected for the resistivity of the wires. This reduction saturates at very low temperatures (around 10 K), since the resistivity of metals is typically a power function of the temperature ($\rho \propto T^5$)[30]. The remaining resistivity at low temperatures originates from defects in the crystal, which cannot be compensated by lowering the temperature. These measurements together with the previous experiments confirm the assumption that the zigzag wires are metallic.

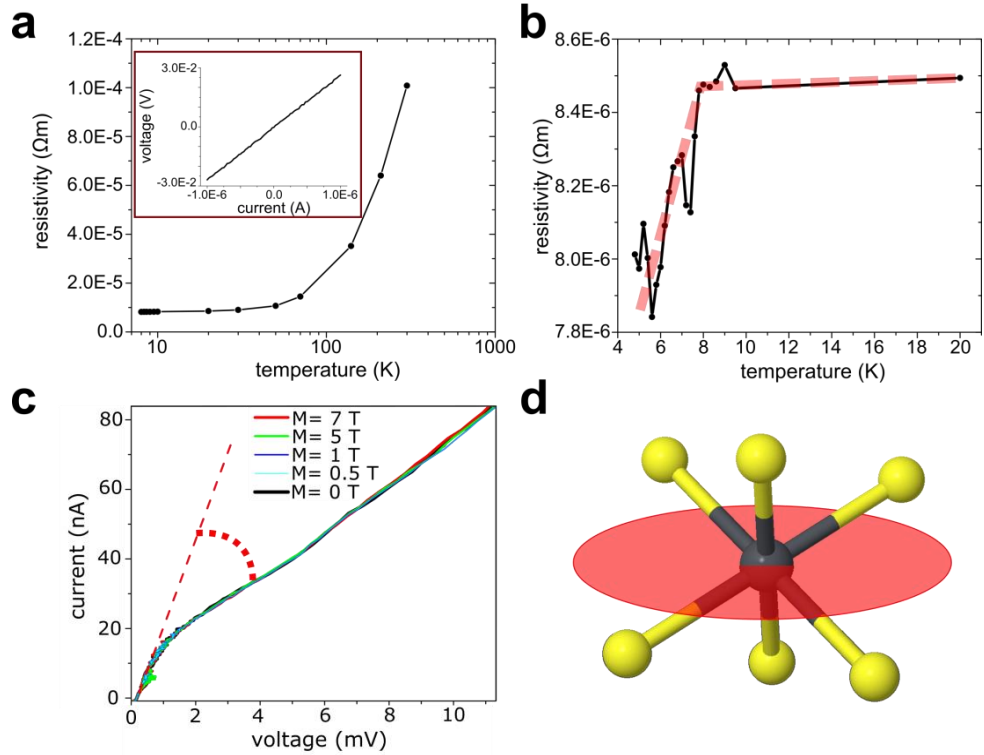

**Figure 5 | Four-point measurements on the zigzag wires.** (**a**) Resistivity of the zigzag wires at different temperatures. The resistivity decreases by reducing the temperature and saturates at very low temperatures. The inset shows the I-V characteristics of the zigzag wires without contact resistance (achieved by four-point measurements). (**b**) Resistivity of the zigzag wires at temperatures below 20 K. A resistivity drop is observed at around 7.5 K, which is considered to be a transition to the superconductivity state. (**c**) I-V characteristics of the wires at 4 K with applied magnetic fields (perpendicular to the substrate) up to 7 T. (**d**) First coordination shell of a lead atom in PbS crystal: Pb shown in grey and S shown in yellow. By cutting the crystal through the {111} facet (the plane shown in red), three of the bonds are broken, leaving the Pb atom with one uncompensated charge.

A surprising property of the zigzag wires is observed at temperatures below 7.5 K. The resistivity of these wires, which is constant below 20 K, experiences a drop below ~ 7.5 K, as shown in Fig. 5b (the four-point conductivity of these wires at different temperatures is demonstrated in Supplementary Figure 8). This temperature is consistent with the transition temperature of bulk Pb into the superconductivity phase[32-35]. Comparable to other Pb nanostructures, the drop of the resistivity occurs with a smooth transition[35, 36]. Therefore, it can be hypothesized that the zigzag wires show features of a transition to the superconductivity state, resulting from the lead terminated surface.

As schematically shown in Fig. 3c (and as it will be discussed in more details in the next section), such a Pb-rich surface includes only one monolayer of Pb. It is known that the 2D superconducting state is extremely vulnerable to quantum phase fluctuations[37, 38]. As a result, the realization of 2D superconductors (based on metals) has remained very difficult and has been achieved only in a few cases[37, 39, 40]. We assume, that the semiconducting body of zigzag wires could also play a crucial role to support the 2D superconducting surface, similar to the previous observations of superconductivity through low dimensional materials supported by semiconducting substrates[37, 38, 41].

The interesting point about ultrathin or 2D superconductors is their good stability in magnetic fields[38, 42, 43]. To test this, we recorded the IV characteristics of these wires while magnetic fields up to 7 T were applied (perpendicular to the substrate) and the temperature was kept at 4 K. As can be seen in Fig. 5c, the drop of the resistance below 20 nA, observable through the significant increase of the slope in the I-V characteristics (which is shown with the red angle in the figure as another representation of the superconductivity) stays identical and the superconductivity is not quenched even at a magnetic field of 7 T, which originates from the marginal thickness of the Pb-rich surface (one monolayer). The fact that the magnetic field cannot be perpendicular to the superconducting plane (due to the special volumetric shape of these wires) amplifies the stability of these wires against magnetic fields, since perpendicular magnetic fields have normally stronger effects to quench the superconductivity[38, 44]. All of these observations emphasize the importance of this new emerging feature of PbS nanostructures, which are normally semiconducting and are modified by faceting the crystal.

The appearance of the superconductivity in lead chalcogenides is rare however a previously reported effect. The superconductive state has been observed in PbS at 6.2 K [45], although we are not aware of the surface conditions of the reported samples. In PbTe, the superconductivity has been reported arising at the interface with different epitaxial metal films such as Pb, In, Tl, Sn in the range of 4–7 K [46]. Recently, Zolotavin and Sionnest [47] studied the transformation of films comprised of colloidally prepared Pb/PbO to the Pb/PbSe core-shell nanocrystals under low temperatures accompanied with a transition to the superconducting state at T<7 K. They found the films to have a finite resistance and 50-fold enhancement of the critical magnetic field compared to the bulk lead. These findings were attributed to peculiarities of granular and disordered systems. In our case, the existence of finite residual resistance might have similar origin based on the fact, that the superconductivity transition occurs at the surface or interface of the PbS (PbS–Pb-monolayer–ligand) and due to the surface/interface interruptions and irregularities might have partially local character within an individual nanowire.

**Discussion**

The performed electrical measurements confirmed that the zigzag wires are metallic. This property originates from the formation of {111} surface facets, which makes the surface Pb-rich. The charge transport mainly occurs through this metallic low resistivity surface (of the zigzag nanowires), in contrast to the straight nanostripes, in which the semiconducting body provides the transport route. In order to clarify why this specific surface facet results in metallic behaviour, a closer look is required on Pb-S bonds of the PbS crystal. As can be seen in Fig. 5d, every Pb atom is bound to 6 S atoms to form the crystal. Inside the regular crystal, Pb donates in total 2 electrons to these S atoms. When the crystal is cut through the {111} facet, half of these bonds (3 of them) are broken and therefore, one electron remains for the Pb atom. These uncompensated charges, which exist on the whole bare surface, change its character to metallic[27].

Although the octahedra, as the building blocks of the zigzag nanowires, are metallic, there are some conditions to observe this metallic behaviour after their attachment (i.e. for the zigzag wires). First of all, the shape of the wires must be well-defined. After the attachment of the octahedra and during the growth phase, the crystal expands in size. In this phase, if any facets other than the {111} facet grow, especially the {100} facets, the shape of the wires would be distorted. This leads to the discontinuity of the metallic path and to the destruction of metallic behaviour (the uncontrolled growth of the wires can be seen in Supplementary Figure 9). Therefore, it is crucial to have a well-defined shape in order to preserve the metallic character. Here, accurately adjusting the synthesis time and employing the right amounts of ligands play a key role to accurately define the crystal surface and to obtain a distinct shape. With our synthesis strategy, zigzag wires are produced with lead terminated and well defined {111} surface facets, compared to previous published synthesis of PbS zigzag wires[17,48].

The other important point is the coverage with Cl$^-$ ions (as the employed X-type ligands) on the surface. As it can be found in the literature and as our DFT calculations show (Supplementary Figure 10a), when the surface of the crystal is completely covered with Cl$^-$ ions (or other ligands), a forbidden gap is formed in the band structure of the crystal[27, 28]. Therefore, semiconducting behaviour must be observed, as the DOS also reaches zero for the bandgap region of such a crystal (Supplementary Figure 10b).

To measure the Cl content on the surface of the wires, they were probed with X-ray photoelectron spectroscopy (XPS)[49, 50]. These measurements (shown in Supplementary Figure 10c and 11) reveal that the Cl content on the surface of the zigzag wires (15%) is lower compared to the straight nanostripes (20 %) and especially compared to the previously studied PbS nanosheets (30-60 %)[51]. In addition, the HAADF-XEDS measurements show a lower Cl content in the whole crystal of the zigzag wires compared to the straight stripes (Supplementary Figure 12). Therefore, it can be concluded that the performed synthesis fulfils this requirement and preserves metallic behaviour of the wires by reducing the Cl content on the surface of the crystal. As possible scenario, TOP present in the synthesis flask could alter the reactivity of the lead precursor and fulfils the requirement of a partial reduction of the superficial lead cations, which further plays a role in the metallicity of the {111} facet. An additional supportive evidence for this give two low-intensity XPS peaks at 136.4 and 141.3 eV which fit to metallic lead and are not present in straight nanostripes. The ratio of $Pb^0$:$PbCl_2$ is 1:2.3 revealing the appearance of combined surface conditions and might lead to residual resistivity. Small relative intensities of the mentioned peaks in comparison to $Pb^{2+}$ in PbS support the argument that the lead surface forms presumably a monolayer.

It is worthy to point out that the metallic nature of the zigzag wires is further evidenced by the width of the XPS signals. Peak narrowing is observed for the core level spectra, compared to the straight nanostripes which might be attributed to the surface metallicity and relative increase in the conductivity. Narrowing of the Full Width at Half Maximum (FWHM) of core levels for more conductive samples has been reported in copper phosphate glasses (O 1s)[52] and carbon blacks (C 1s)[53] and attributed to structural changes which can be, to some extent, correlated to an increase in the conductivity.

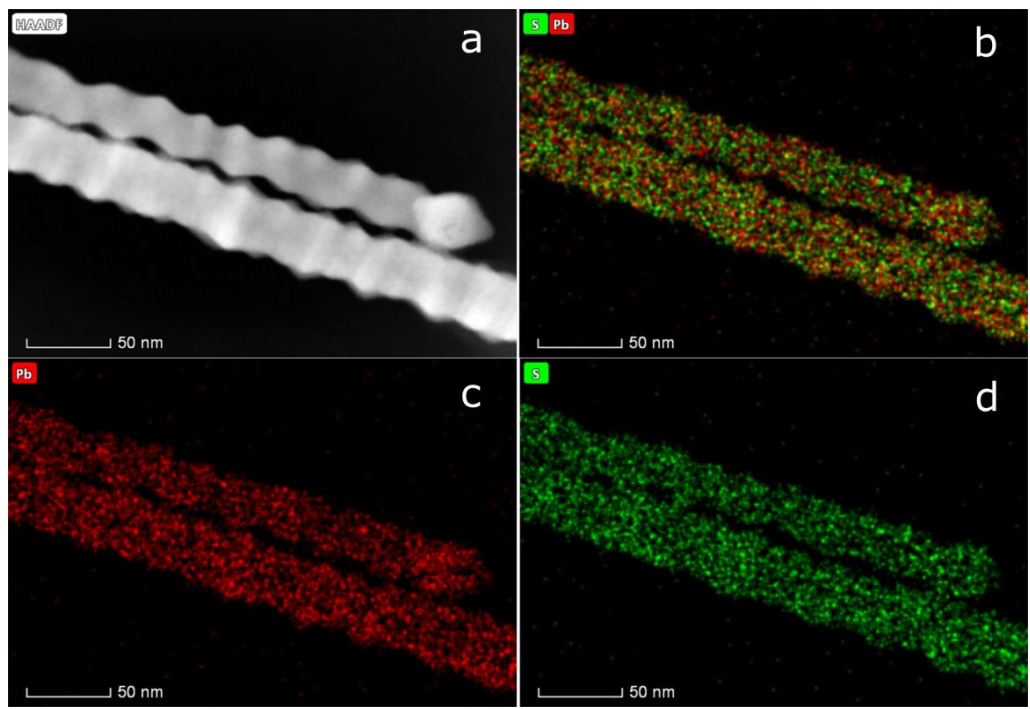

**Figure 6 | High-angle annular dark-field imaging (HAADF) of zigzag wires.** (**a**) An HAADF image of exemplary zigzag wires (**b**) HAADF-XEDS element mapping of zigzag PbS nanowires showing a homogenous distribution of (**c**) lead and (**d**) sulphur atoms without any hints of an existing massive lead shell grown around the wires.

One might argue that the Pb-terminated surface of the zigzag wires could be a foundation for the growth of a pure Pb shell around the wires, which could also render them metallic. To address this question, the HRTEM image of these wires (shown in Fig. 2c) was analysed. As can be seen, there is no distortion of the crystal at the surface and its periodicity is constant. HAADF-XEDS element mapping was also performed on zigzag nanowires and shows a homogenous distribution of lead and sulphur (shown in Fig. 6). From XEDS we find that the Pb/S ratio for the zig-zag wires is in the range of 1.15–1.2, which slightly deviates from the stoichiometry of truncated Pb-terminated octahedra with a base of 30 nm (Pb/S ≈ 1.1). Taking into account the EDX measurement errors for PbS, we note a rough agreement with DFT simulations[54] of PbS quantum dots where metallic behaviour was induced based on the off-stoichiometry. For the flat nanostripes, Pb/S≈1.16–1.18, and increase of chlorine amount was observed. Since Pb Mα and S Kα, Kβ interfere dramatically, the atomic errors for this kind of measurement are relatively large (6.2% for S and 9.5% for Pb) with fitting errors of 0.9 and 0.3 %. Further, XRD measurements do not show any sign of massive lead. This demonstrates that the whole crystal is PbS or in other words, no volumetric Pb shell is formed around the wires. This could also be confirmed by XPS measurements as shown in Supplementary Figure 10c. They demonstrate that massive Pb species are absent from both samples, since their presence could be inferred by the observation of a sharp, asymmetric peak, observable at low binding energies[55].

**Conclusion**

The electrical properties of PbS nanowires were tuned from semiconducting to metallic and supposedly to superconducting by faceting the crystal in a distinct way. While various forms of PbS nanostructures are semiconducting, when the crystal is cut through the {111} facets, a Pb-terminated {111} surface around the crystal renders it metallic. Such a crystal was formed by colloidal chemistry by accurately controlling the growth. Several methods including DFT calculations, field-effect, photoconductivity, and temperature dependency measurements were employed to confirm the metallic character of these wires. It turned out that the well-defined shape along with the low amounts of Cl content and the delocalized character of surface states play crucial roles for preserving metallic behaviour. These results represent a new exciting platform to further optimize the colloidal materials and to predictably tune their properties according to the target application. Further, they might open new pathways for realizing superconducting monolayers based on colloidal crystalline materials.

**Materials and methods**

All chemicals were used as received. The following chemicals were used: diphenylether (Sigma-Aldrich, 99%), lead(II)acetate trihydrate (Sigma-Aldrich, 99.999%), lithium chloride (Sigma-Aldrich, >=99%), N,N-dimethylformamide (DMF, Sigma-Aldrich, 99.8% anhydrous), oleic acid (OA, Sigma-Aldrich, 90%), thiourea (TU, Alfa Aeser, >=99% ), toluene (VWR, >=99.5%), tri-*n*-octylphosphine (TOP, abcr, 97%).

*Synthesis of PbS Stripes*

Different types of lead sulphide nanowires were synthesized as followed. In a three-neck-flask with thermocouple, condenser and septum 860 mg $PbO(Ac_2) \cdot 3H_2O$ (2.3 mmol), 3.5 mL OA (10 mmol) and 10 mL Diphenylether are mixed and degassed at 70 °C in vacuum for at least 2.5 h. Under nitrogen atmosphere, 0.10 mL TOP (0.22 mmol) is added and the solution is heated up to 200 °C. While heating, 0.2 mL of a 0.3 M LiCl-DMF solution is added. When the reaction solution reaches the desired temperature, 0.2 mL of a 0.09 M TU-DMF solution is added. Then, the solution turns black, shows metallic luster and is stirred for 90 s until the heating mantle is removed. When the reaction solution reaches room temperature, it is centrifuged with 7 k rpm (6695 rcf) for 5 min. The colourless supernatant is disposed and the black precipitate is washed two more times by dispersion in toluene and centrifugation as described before. The wires, which are stable for at least some weeks, are stored in toluene.

*TEM and SEM characterization*

Transmission electron microscope (TEM) images and selected area electron diffraction (SAED) were performed with a JEOL-1011 transmission electron microscope (Jeol, Tokyo, Japan, at 100 kV). The high resolution (HR) TEM and HAADF-STEM images were obtained by using a Cs corrected (TEM and STEM, CEOS) JEOL JEM-2200 FS (200 kV). SEM images were obtained from a FEI Quanta 3D FEG microscope.

*DFT calculations*

The density function theory package ABINIT package[56, 57], a common project of the Université Catholique de Louvain, Corning Incorporated, and other contributors (URL http://www.abinit.org) was used for band structure and density of states calculations. As exchange functional LDA[58] and Hartwigsen-Goedecker-Hutter pseudopotentials[59] were used. The lattice parameters were fixed to the experimental values. In confinement, an additional vacuum of at least 10 Angstrom was added to the slab in c direction.

*Device preparation and measurements*

A diluted suspension of the wires was spin-coated on a $Si/SiO_2$ substrate, and contacted individually by electron-beam lithography, followed by thermal evaporation of 1/50 nm Ti/Au. The achieved devices were transferred to a probe station for room or low temperature electrical measurements in vacuum. All the measurements were carried out with back-gate geometry, using a highly doped silicon substrate with 300 nm thermal oxide as gate dielectric. For the illumination of the wires, a red laser ($\lambda = 630$ nm) with 20 mW power was used. In order to probe the reliability of the results, every measurement has been performed/repeated on several devices which all had comparable results. The results shown in the manuscript are the best representing outcomes.

*XPS measurements*

X-ray photoelectron spectroscopy (XPS) measurements were carried out using a high-resolution 2D delay line detector. A monochromatic Al K α X-ray source (photon energy 1486.6 eV, anode operating at 15 kV) was used as incident radiation and, to compensate for charging effects, a flood gun was used. XPS spectra were recorded in fixed transmission mode. A pass energy of 20 eV was chosen, resulting in an overall energy resolution better than 0.4 eV. The binding energies were calibrated based on the graphitic carbon 1 s peak at 284.8 eV[60].

*Tomography and EDS mapping*

HAADF-STEM XEDS maps and Tomography experiments were performed using a double aberration-corrected FEI Titan$^3$ Themis 60–300 microscope equipped with 4-detector ChemiStem system. Very high spatial resolution STEM-XEDS maps were acquired using a high brightness, subangstrom (0.07 nm) diameter, electron probe in combination with a highly stable stage which minimized sample drift. Element maps were acquired with a screen current of 60–100 pA and a pixel time of 0.2 ms which results in a total acquisition time of approximately 30 minutes. An averaging filter was used on the images as provided in the Velox software from FEI. To acquire the STEM tomography tilt series, a convergence angle of 9 mrad was selected in order to improve the depth of focus and a camera length of 115 mm was used. The software FEI Explore3D v.4.1.2 enabled the acquisition of the tomography tilt series from −62° to +62° every 2° and the alignment and reconstruction of the data set. Avizo software was used for visualization.

**Acknowledgments**: M.M.R.M., R.L. and C.K. gratefully acknowledge financial support of the European Research Council via the ERC Starting Grant "2D-SYNETRA" (Seventh Framework Program FP7, Project: 304980). C.K. thanks the German Research Foundation DFG for financial support in the frame of the Heisenberg scholarship KL 1453/9-2. M.M.R.M. thanks the PIER Helmholtz Graduate School for the financial support. A.B.H. thanks MINECO for the project MAT2016-81118-P.

**Competing financial interests**: The authors declare no competing financial interests.